\documentclass{moriond}

\def\be{\begin{equation}}
\def\ee{\end{equation}}
\def\bea{\begin{eqnarray}}
\def\eea{\end{eqnarray}}
\newcommand{\gsim}{\raisebox{-0.13cm}{~\shortstack{$>$ \\[-0.07cm]
			$\sim$}}~}
\newcommand{\lsim}{\raisebox{-0.13cm}{~\shortstack{$<$ \\[-0.07cm]
			$\sim$}}~}

\newcommand{\Photo}{\includegraphics[height=35mm]{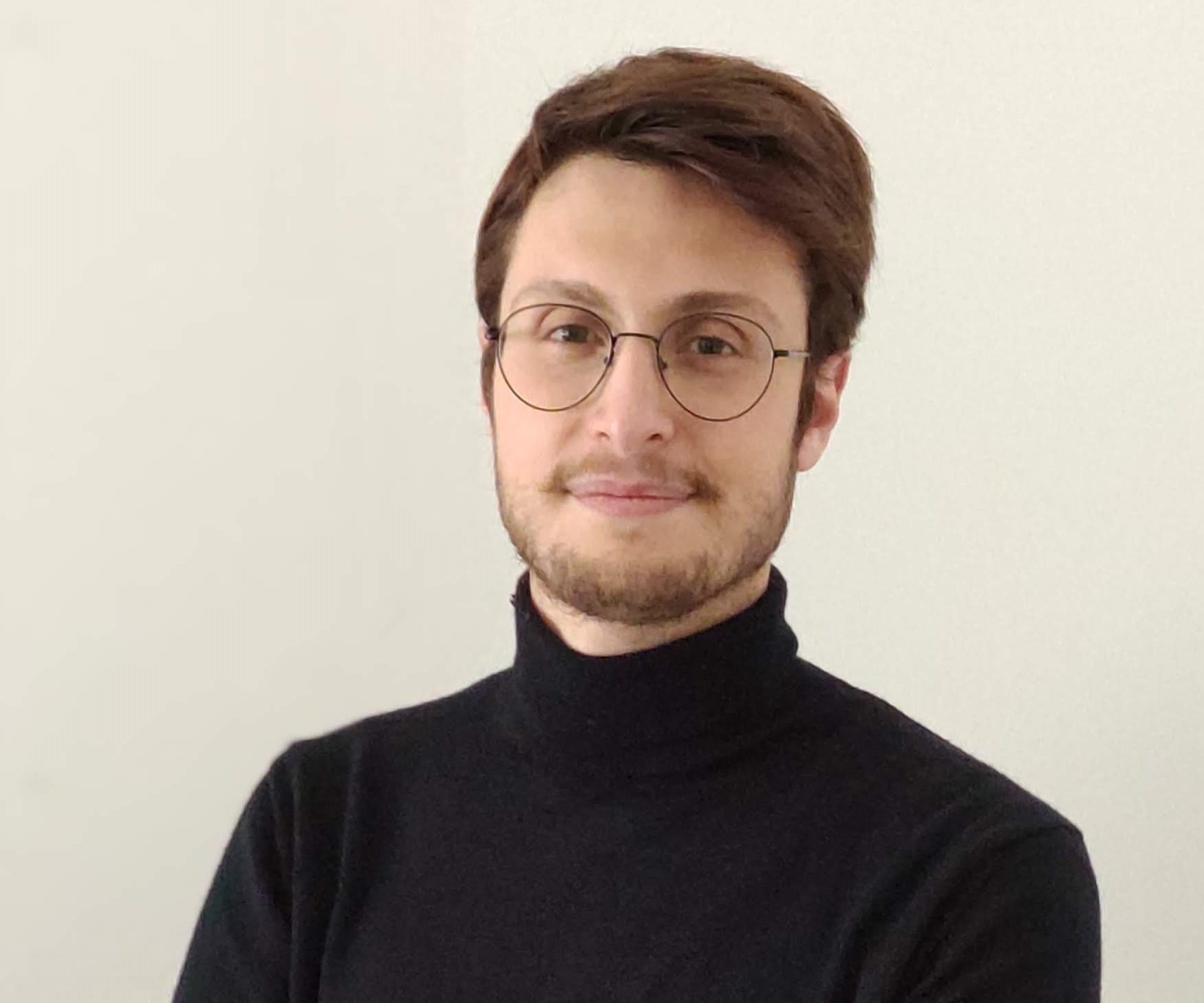}}

\begin{document}
\vspace*{4cm}
\title{SPONTANEOUS CP VIOLATION}

\author{ALESSANDRO VALENTI}

\address{Dipartamento di Fisica e Astronomia ``G.~Galilei", Università di Padova, Italy}

\maketitle\abstracts{
Models of spontaneous CP violation can solve the Strong CP problem without the need of an anomalous Peccei-Quinn symmetry. In this work we review the Nelson-Barr approach, quantifying a peculiar coincidence between unrelated mass scales that these models must satisfy in order to correctly reproduce the Standard Model quark masses and CP violation. We investigate the compatibility between this requirement and the induced radiative corrections to the neutron electric dipole moment, and with bounds coming from collider, electroweak and flavor observables.}

\section{Introduction}

The colored sector of the Standard Model (SM) features two fundamental sources of CP violation. The first is the irreducible phase of the CKM matrix, which shows up in flavor-violating observables and whose experimental value is of order one, $\delta _{\mbox{\footnotesize CKM}} \simeq 1.2$ \cite{Zyla:2020zbs}. The second one is the coefficient of the QCD topological term $ ({\footnotesize g^2 _s / 32\pi^2})~G\tilde G$, given by $\bar \theta = \theta_{QCD}+\arg \det Y_u Y_d$. This angle is relevant only in the strongly coupled phase of QCD and can be related to the neutron, atoms and molecules electric dipole moments, whose measurements~\cite{Abel:2020pzs} set the bound $\bar \theta \lsim 10^{-10}$. The Strong CP problem can then be formulated as a fine-tuning problem of ten orders of magnitude between two parameters sharing a common origin, the SM Yukawas.
Many solutions have been proposed to solve this puzzle, relying on anomalous continuos symmetries~\cite{Peccei:1977hh} or on the discrete symmetries P~\cite{Beg:1978mt} or CP~\cite{Georgi:1978xz}. Since P and CP can be exact symmetries of the UV lagrangian, these solutions are not subject to the problem of the quality of the underlying symmetry principle. As such P or CP models are fundamentally different from the QCD axion solution, where this might become an issue~\cite{Kamionkowski:1992mf}. In this work we analyse CP scenarios which fall in the class of Nelson-Barr models, in which the spontaneous breaking of CP, ultimately needed in order to reproduce $\delta_{\mbox{\footnotesize CKM}}$, is obtained in a hidden sector and mediated to the SM through a fermionic messenger. In particular we focus on a simple but effective model~\cite{Bento:1991ez}, dubbed hereafter as $d-$mediation. For a more general analysis we refer to other works \cite{Vecchi:2014hpa,Dine:2015jga,Valenti:2021rdu}.

\section{The Nelson-Barr $d-$mediation}
In the minimal model of $d-$mediation the down-quark Yukawa sector of the SM lagrangian is modified by the addition of a pair of complex scalar singlets $\Sigma$  and a fermionic mediator $\psi$ as 
\be
-\mathcal{L}_{\mbox{\footnotesize Yuk}} ^d = y_d \,\bar Q H d + y\,\bar \psi \Sigma d + m_{\psi} \, \bar \psi \psi
\label{eq:lagr1}
\ee
where $\psi, \Sigma$ are assumed to be charged under some $U(1)$ to avoid additional couplings with other SM fields. Due to CP invariance all the couplings and the masses can be taken as real. The field $\Sigma$ breaks CP spontaneously by acquiring a complex vacuum expectation value, such that $\xi \equiv y \langle \Sigma \rangle$ is the effective order parameter of CP violation from the SM perspective
. After CP breaking the lagrangian (\ref{eq:lagr1}) can be put in the following form by means of a flavor $SU(4)$ rotation of $(d,\psi)$:
\begin{equation}
	-\mathcal{L}_{\mbox{\footnotesize Yuk}} ^d = Y_d \, \bar Q H d + Y\, \bar Q H \psi + M\, \bar \psi \psi \;  , \qquad \qquad \left\{
	\begin{array}{l}
		Y_d = y_d \left[1-\frac{\xi\xi^{\dagger}}{|\xi|^2}\left(1-\frac{m_{\psi}}{M}\right) \right] \\
		M = (|\xi|^2 + m_{\psi} ^2 )^{1/2}\\
		Y= y_d \frac{\xi}{M} = Y_d \frac{\xi}{m_{\psi}}
	\end{array}
\right. .
\label{eq:lagr2}
\end{equation}
From the explicit form of the lagrangian above it is easy to check that the determinant of the four-dimensional mass matrix is real, so that $\bar \theta_{\mbox{\footnotesize tree}}=0$, while at the same time the physical Yukawa $Y_d$ has inherited CP violation by becoming complex.

\subsection{Quark masses and the CKM phase}
\label{sec:2.1}
After the $SU(4)$ rotation, the physical Yukawa $Y_d$ in (\ref{eq:lagr2}) is given by a particular combination of $y_d$ and $\xi$. The first thing to notice is that $\det Y_d = \det y_d \, (m_{\psi}/M)$, so that $m_{\psi} \ll |\xi|$ is forbidden if one wishes to obtain a realistic mass spectrum. The parameters $m_{\psi},\,|\xi|$ are further constrained by the requirement of reproducing the CKM phase. In the basis where $Y_u$ is diagonal, the CKM matrix diagonalises the combination $Y_d Y_d^{\dagger} = y_d (1-\frac{\xi \xi^{\dagger}}{M^2}) y_d ^t$, which has $\mathcal{O}(1)$ complex entries if $|\xi| \sim m_{\psi}$. A careful analysis~\cite{Valenti:2021rdu} reveals that $\delta_{\mbox{\footnotesize CKM}} \sim 1.2$ only for $|\xi|/m_{\psi }\gsim2$. At the same time the coupling $Y$ in (\ref{eq:lagr2}) is constrained by the requirement of perturbativity to be $\ll 4\pi$, which in turn forces $|\xi|/m_{\psi }\ll 4\pi /y_b$. In this way we can summarize the viable window for the parameters as
\begin{equation}
	2 \lsim \frac{|\xi| }{m_{\psi}} \ll 10^3.
	\label{eq:ckmwindow}
\end{equation}
This puzzling coincidence of CP-even and CP-odd mass scales can not be addressed in an effective framework, but requires an explicit UV construction of the CP breaking sector.

\subsection{Radiative corrections to $\bar \theta$}
\label{sec:2.2}
Given the strength of the experimental constraint on $\bar \theta$, it is extremely important to check that direct (to the topological parameter) and indirect (to the quark masses) radiative corrections are under control. At 1-loop, by diagrammatic inspection it is clear that these must involve the scalar mixing $\lambda_{ab} |H|^2 \Sigma ^{\dagger} _a \Sigma _b$, with the external $\Sigma$ and $H$ set on their vev. This mixing is related to the hierarchies of the scalar sector (i.e. the Naturalness problem) and thus its size is model-dependent, even though its contribution to $\bar \theta$ is typically under control~\cite{Bento:1991ez}. At 2- and 3-loops there are contributions which involve only the SM quarks and the fermionic mediator $\psi$. Interestingly, these are directly related to the mediation of CP violation to the SM and thus are \emph{irreducible}~\cite{Valenti:2021rdu}. In this class a distinction can be made between non-analytical and analytical contributions. The former involve logs of the physical quark masses  $m_q$, are suppressed by powers of $(m_{q}/M)$ and are under control for $M \gsim$TeV. The analytical ones involve only analytical functions of the Yukawas. In this case the only appearing mass ratio is $|\xi|/m_{\psi}$, which is already constrained by the reproduction of the physical CKM phase (\ref{eq:ckmwindow}). The compatibility of this constraint with the bounds from $\bar \theta$ is non-obvious and a crucial aspect of these models. For one family of mediators, the first irreducible contribution appears at three loops and results in $\bar \theta \sim (1/16\pi^2)^3\, \mbox{Im} \, Y^{\dagger} (Y_d Y_d ^{\dagger} Y_u Y_u ^{\dagger}) Y$. This can be evaluated plugging the expression in (\ref{eq:lagr2}) for $Y$ and the numerical values of $Y_u,Y_d$ in the physical basis, and gives a milder bound on $|\xi|/m_{\psi}$ with respect to (\ref{eq:ckmwindow}). For more families of mediators, flavor violation in the mediator sector opens the possibility for more CP-violating structures which give stronger constraints: an example is $\bar \theta \sim  (1/16\pi^2)^3\, \mbox{Im} \, \mbox{tr} (Y^{\dagger} Y_u Y_u ^{\dagger} Y Y^{\dagger} Y F(M^{\dagger} M))$, where $M$ and $Y$ are the generalizations of the quantities in (\ref{eq:lagr2}) to more families of mediators and $F$ a generic dimensionless function of the mediators' masses. Inserting the experimental values, this contribution results in~\footnote{\label{footnote}~For the sake of clarity, here and in the following we employ a slight abuse of notation by keeping on calling $|\xi|/m_{\psi}$ the relevant parameter for CP violation even in the case of more families of mediators. More correctly we should refer to $Y_d ^{-1} Y$, which is of order $|\xi_{ij}|/|m_{\psi, ij}|$ if no strong hierarchies or accidental cancellations in the mediator sector are present. For a more detailed treatment see ref. \cite{Valenti:2021rdu}.} $\bar \theta \sim 6 \times 10^{-18} (|\xi|/m_{\psi})^4$ which gives an upper bound on $|\xi|/m_{\psi}$ stronger than (\ref{eq:lagr2}).

\subsection{Constraints on the mediator}
\label{sec:2.3}
The mediator $\psi$ is a vector-like copy of the down-quarks, and as such a pletora of direct and indirect constraints are available~\cite{CMS:2020ttz}. Direct searches set the bound $M\gsim 1.4$~TeV. Low-energy flavor and CP observables lead to constraints on the Yukawa interaction $Y \, \bar Q H \psi$, translating in turn to bounds on the mediation of CP violation. Since $Y\sim Y_d (\xi/m_{\psi})$, the most important constraints come from processes involving the heaviest down quarks. At tree-level the new Yukawa modifies the coupling of the $b$ quarks to the $Z$ boson, accurately measured at LEP. Constraints from $\Delta F=2$ transitions are described in terms of four-fermions operators involving left-handed down-quarks, which are induced both at tree-level, suppressed by $(v/M)^2$, and at 1-loop through box diagrams with virtual $H$ and $\psi$. Due to the hierarchies in $Y$ the most important bound comes from $B_s ^0-\bar B_s ^0$ oscillations. $\Delta F=1$ transitions are captured by semileptonic four-fermions operators induced at tree-level and are bounded by rare $B_s ^0$ meson decays as $B_s ^0 \rightarrow \bar \ell \ell, X^0 _s \bar \ell \ell$. Other $\Delta F=1$ observables (e.g. $B^0_s \rightarrow X^0_s \gamma$) are induced at 1-loop and are subleading.

\begin{figure}[!h]
	\begin{minipage}{0.45\linewidth}
		\centerline{\includegraphics[width=\linewidth]{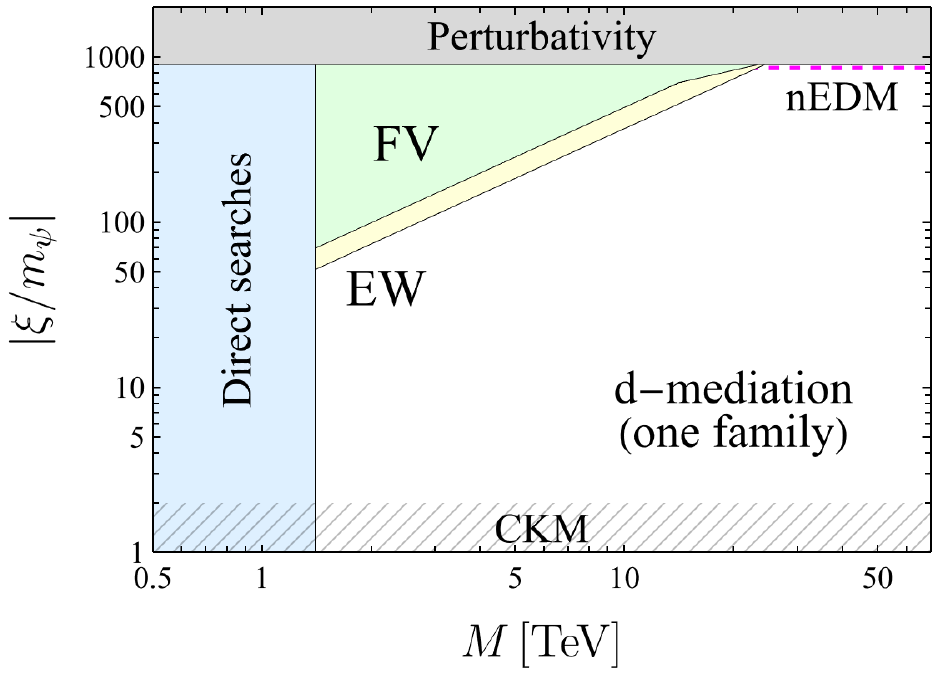}}
	\end{minipage}
	\hfill
	\begin{minipage}{0.45\linewidth}
		\centerline{\includegraphics[width=\linewidth]{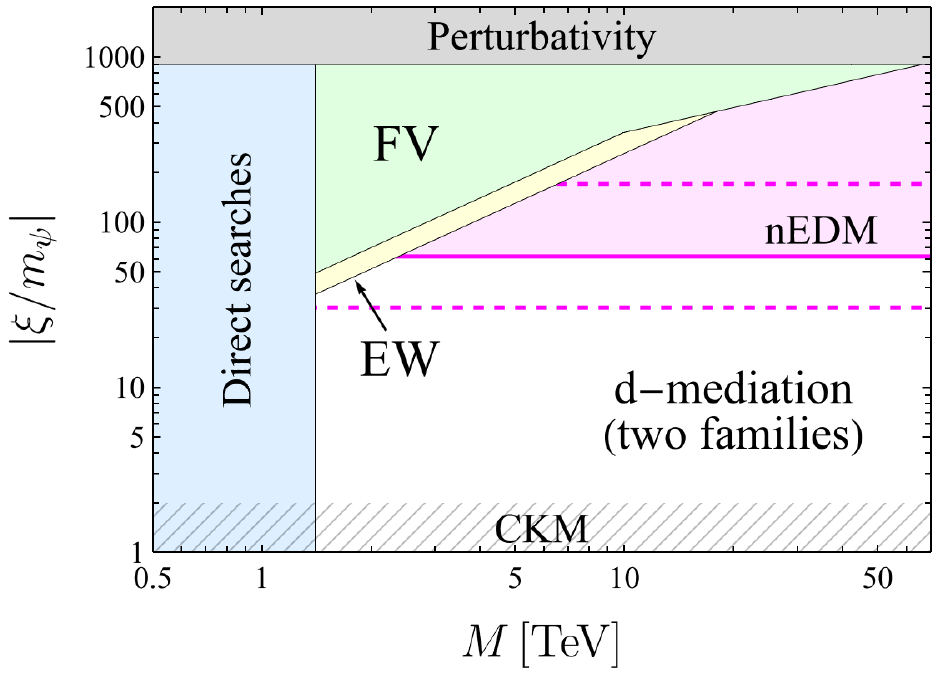}}
	\end{minipage}
	\caption{Summary of the constraints discussed in section \ref{sec:2.1} (CKM, Perturbativity), \ref{sec:2.2} (nEDM) and \ref{sec:2.3} (Direct searches, EW and FV) for the case of one and two families of mediators. In the latter case we assume $M_1=M_2$  and no hierarchies in the mediators' sector, as discussed in footnote~$^{\ref{footnote}}$.}
	\label{fig:constraints}
\end{figure}

In figure \ref{fig:constraints} we summarize the constraints discussed in section \ref{sec:2.1}, \ref{sec:2.2} and \ref{sec:2.3} for the case of one and two families of mediators. While the lower bound on $|\xi|/m_{\psi}$ is set by the CKM phase, the interplay between direct and indirect observables has just started to fill the upper part of the plot, leaving a large portion of the parameter space still available. For more families of mediators the most important upper bound comes by the non-observation of the neutron electric dipole moment.  

\section{Conclusion}
The Strong CP problem is one of the most interesting challenges for BSM physics. In this work we reviewed the Spontaneous CP violation approach, specialising to Nelson-Barr $d-$mediation models. In these scenarios the requirement of reproducing the quark masses and the CKM phase forces the non-trivial coincidence $2 \lsim |\xi|/m_{\psi} \ll 10^3$. This viability range must be confronted with bounds coming from radiative corrections to $\bar \theta$ and from the new Yukawa coupling $Y$, both regulated by the very same ratio $|\xi|/m_{\psi}$. While in models of $d-$mediation these constraints are typically compatible, as shown in fig. \ref{fig:constraints}, this is not true for the cases of $q-$ and $u-$mediation (for more families of mediators), where corrections to $\bar \theta$ are unacceptably large~\cite{Vecchi:2014hpa,Valenti:2021rdu}.

The peculiar coincidence between a priori unrelated CP-even ($m_{\psi}$) and CP-odd ($\xi$) parameters must be addressed in any completion of this effective framework if one wishes to really solve the Strong CP problem without barely transmuting it into another hierarchy problem. Explicit UV constructions which succeed in facing this issue require either supersymmetry~\cite{Hiller:2001qg} or a strongly-coupled sector~\cite{Valenti:2021xjp}. Indeed, in four dimensions these are virtually the unique options able to generate non-trivial coincidences of mass scales in a natural way.

\section*{Acknowledgments}
The author thanks L.Vecchi for his collaboration on the project. This project has received support from the European
Union’s Horizon 2020 research and innovation programme under the Marie Sklodowska-Curie grant
agreement No 860881-HIDDeN.

\end{document}